\documentclass[12pt]{article}
\usepackage{amssymb}
\def\hybrid{\topmargin 0pt      \oddsidemargin 0pt
        \headheight 0pt \headsep 0pt
        \voffset=-0.5cm
        \hoffset=-0.25in
        \textwidth 6.75in
        \textheight 9.5in       % A4 paper
        \marginparwidth 0.0in
        \parskip 5pt plus 1pt   \jot = 1.5ex}
\catcode`\@=11
\def\marginnote#1{}

\newcount\hour
\newcount\minute
\newtoks\amorpm
\hour=\time\divide\hour by60 \minute=\time{\multiply\hour by60
\global\advance\minute by-\hour}
\edef\standardtime{{\ifnum\hour<12 \global\amorpm={am}%
        \else\global\amorpm={pm}\advance\hour by-12 \fi
        \ifnum\hour=0 \hour=12 \fi
        \number\hour:\ifnum\minute<10 0\fi\number\minute\the\amorpm}}
\edef\militarytime{\number\hour:\ifnum\minute<10 0\fi\number\minute}

\def\draftlabel#1{{\@bsphack\if@filesw {\let\thepage\relax
   \xdef\@gtempa{\write\@auxout{\string
      \newlabel{#1}{{\@currentlabel}{\thepage}}}}}\@gtempa
   \if@nobreak \ifvmode\nobreak\fi\fi\fi\@esphack}
        \gdef\@eqnlabel{#1}}
\def\@eqnlabel{}
\def\@vacuum{}
\def\draftmarginnote#1{\marginpar{\raggedright\scriptsize\tt#1}}
\def\draftlabel#1{{\@bsphack\if@filesw {\let\thepage\relax
   \xdef\@gtempa{\write\@auxout{\string
      \newlabel{#1}{{\@currentlabel}{\thepage}}}}}\@gtempa
   \if@nobreak \ifvmode\nobreak\fi\fi\fi\@esphack}
        \gdef\@eqnlabel{#1}}
\def\@eqnlabel{}
\def\@vacuum{}
\def\draftmarginnote#1{\marginpar{\raggedright\scriptsize\tt#1}}

\def\draft{\oddsidemargin -.5truein
        \def\@oddfoot{\sl preliminary draft \hfil
        \rm\thepage\hfil\sl\today\quad\militarytime}
        \let\@evenfoot\@oddfoot \overfullrule 3pt
        \let\label=\draftlabel
        \let\marginnote=\draftmarginnote
   \def\@eqnnum{(\theequation)\rlap{\kern\marginparsep\tt\@eqnlabel}%
\global\let\@eqnlabel\@vacuum}  }

\def\numberbysection{\@addtoreset{equation}{section}
        \def\theequation{\thesection.\arabic{equation}}}

\def\underline#1{\relax\ifmmode\@@underline#1\else
        $\@@underline{\hbox{#1}}$\relax\fi}

\def\titlepage{\@restonecolfalse\if@twocolumn\@restonecoltrue\onecolumn
     \else \newpage \fi \thispagestyle{empty}\c@page\z@
        \def\thefootnote{\fnsymbol{footnote}} }

\def\endtitlepage{\if@restonecol\twocolumn \else  \fi
        \def\thefootnote{\arabic{footnote}}
        \setcounter{footnote}{0}}  %\c@footnote\z@ }
\numberbysection \hybrid

\newcounter{mo}

\newcommand{\tr}{{\rm tr}}
\newcommand{\ti}[1]{\tilde{#1}}

\newcommand{\mL}{{\mathcal L}}

\newcommand{\mH}{{\mathcal H}}

\newcommand{\la}{\lambda}

\newcommand{\al}{\alpha}
\newcommand{\be}{\beta}
\newcommand{\ga}{\gamma}
\newcommand{\om}{\omega}

\newcommand{\Mat}{ {\rm Mat}(N,\mathbb C) }
\newcommand{\MatM}{ {\rm Mat}(M,\mathbb C) }

\newcommand{\mC}{\mathbb C}
\newcommand{\mZ}{\mathbb Z}

\newcommand{\eps}{\varepsilon}
\def\beq{\begin{equation}}
\def\eq{\end{equation}}
\def\p{\partial}

\newcommand{\mats}[4]{\left(\begin{array}{cc}{#1}&{#2}\\ {#3}&{#4}
\end{array}\right)}

\begin{document}

\setcounter{page}{1}

\date{}
\date{}
%\vspace{30mm}

\begin{flushright}
ITEP-TH-11/20
\end{flushright}
\vspace{0mm}

\begin{center}
\vspace{0mm}
 {\LARGE{Quantum-classical correspondence }}
\\ \vspace{4mm} {\LARGE{for ${\rm gl}(1|1)$ supersymmetric Gaudin magnet}}
\\ \vspace{4mm} {\LARGE{with boundary}}
\\
\vspace{15mm} {\large  {M. Vasilyev}
{\small $^{ \diamondsuit\, \flat\, \S}$}
 \ \ \ \ \ \ \ {A. Zabrodin}
 {\small $^{ \flat\, \sharp\, \ddagger}$}
 \ \ \ \ \ \ \ {A. Zotov}
 {\small $^{\diamondsuit\, \natural\, \ddagger\, \S}$}
  }
 \vspace{10mm}

\vspace{1mm} $^\diamondsuit$ --
{\small{\rm
 Steklov Mathematical Institute of Russian Academy of Sciences,\\ Gubkina str. 8, Moscow,
119991,  Russia}}
 \\
 \vspace{1mm} $^\flat$ --
 {\small{\rm Skolkovo Institute of Science and Technology,
 Nobel str. 1, Moscow,  143026 Russia}}
\\
 \vspace{1mm}$^\S$ - {\small{\rm National Research University Higher School of Economics, 
 % \\
%  Mathematics Department of
%  Myasnitskaya str. 20,
Moscow, 
%101000,
 Russia}}
  \\
 \vspace{1mm} $^\sharp$ --
 {\small{\rm Institute of
Biochemical Physics, Kosygina str. 4, 119334, Moscow, Russia}}
\\
 \vspace{1mm} $^\ddagger$ -- {\small{\rm %Institute of Theoretical and Experimental Physics, 117218,  Moscow, Russia
 ITEP NRC KI, B. Cheremushkinskaya str. 25,  Moscow, 117218, Russia}}
\\
  \vspace{1mm} $^\natural$ -- {\small{\rm Moscow Institute of Physics and Technology,\\
  Inststitutskii per. 9, Dolgoprudny,
 Moscow region, 141700, Russia}}

\end{center}

%\vspace{2mm}
\begin{center}\footnotesize{{\rm E-mails:}{\rm\
 mikhail.vasilyev@phystech.edu,\ zabrodin@itep.ru,\ zotov@mi-ras.ru}}\end{center}
%
%\vspace{0mm}
%

 \begin{abstract}
We extend duality between the quantum integrable Gaudin models with boundary
and the classical Calogero-Moser systems associated with root systems of
classical Lie algebras $B_N$, $C_N$, $D_N$ to the case of
supersymmetric ${\rm gl}(m|n)$  Gaudin models with $m+n=2$.
Namely, we show that the spectra of quantum Hamiltonians for all
such magnets being identified with the classical particles velocities provide the zero level of the classical action variables.
 \end{abstract}

%\bigskip

%\bigskip

\newpage
\tableofcontents

 \section{Introduction: an overview}
\setcounter{equation}{0}

 \paragraph{KZ equations and many-body systems.}
In this paper we study the quantum-classical duality appeared previously
 in a number of different contexts \cite{Givental,MTV,Alex,NRS,GZZ,TsuboiZZ,VZZ}. In the general case it is
a certain relation between classical integrable many-body systems and quantum spin chains
 or Gaudin models.
In its simplest form the duality relation follows from the quasiclassical limit
of the Matsuo-Cherednik projection \cite{KZ,MaCh,ReV}.
Namely, consider the ${\rm gl(2)}$ Knizhnik-Zamolodchikov equations
 \beq\label{q01}
 \begin{array}{c}
 \displaystyle{
 \kappa \p_{z_i}\Psi={\bf H}_i^{\hbox{\tiny{G}}}
 \Psi\,,\quad\Psi\in{\mathcal H}\,,\quad i=1\,, \ldots \,,N\,,
 }
 \end{array}
  \eq
where the operators ${\bf H}_i^{\hbox{\tiny{G}}}$ are the Gaudin Hamiltonians \cite{Gaudin}
\beq\label{q02}
 \begin{array}{c}
  \displaystyle{
 {\bf H}_i^{\hbox{\tiny{G}}}={\bf w}^{(i)}+\hbar\sum\limits_{k\neq i}
 \frac{{\bf P}_{ik}}{z_i-z_k}\,, \quad i=1, \ldots ,N
 }
 \end{array}
  \eq
 acting on the Hilbert space ${\mathcal H}=(\mC^2)^{\otimes N}$, ${\bf P}_{ik}$ are permutation
 operators exchanging $i$-th and $k$-th tensor components of ${\mathcal H}$, $\hbar$
 and $\kappa$ are constant parameters (in general complex), and
 ${\bf w}^{(i)}$  acts as constant diagonal (twist) matrix ${\rm diag}(\om,-\om)$
 in the $i$-th component of ${\mathcal H}$.

The Matsuo-Cherednik construction provides a symmetrized
 projection $\langle\Omega|\Psi\rangle$
 of the solution $\Psi$ (\ref{q01}) to a solution
 of the eigenvalue problem for the quantum Calogero-Moser $N$-body system \cite{Calogero}:
\beq\label{q03}
 \begin{array}{c}
  \displaystyle{
 \Big(-\frac{\kappa^2}{2}\sum\limits_{i=1}^N\p_{z_i}^2
 +(\hbar - \kappa)\hbar\sum\limits_{i< j}^N \frac{1}{(z_i-z_j)^2} \Big)
 \langle\Omega|\Psi\rangle=E\langle\Omega|\Psi\rangle\,,
 }
 \end{array}
  \eq
 where the eigenvalue $E$ is a function of the twist parameter $\om$.
 The dual vector $\langle\Omega|\in\mH^*$ is invariant
 with respect to the action of permutation operators.
 Details and generalizations can be found in \cite{FV,GrZZ}.

The quasiclassical limit $\kappa \rightarrow 0$ of the Knizhnik-Zamolodchikov equations (\ref{q01}),
with $\Psi$ expanded as
$\Psi=(\Psi_0+\kappa \Psi_1+ \ldots )e^{S/\kappa }$, with some function
$S=S(z_1, \ldots , z_N)$ leads to the eigenvalue problems
 \beq\label{q04}
 \begin{array}{c}
  \displaystyle{
 {\bf H}_i^{\hbox{\tiny{G}}}\psi=H_i^{\hbox{\tiny{G}}}\psi\,,\quad H_i^{\hbox{\tiny{G}}}=\p_{z_i}S,
 \quad \psi=\Psi_0\in{\mathcal H}\,,\quad i=1, \ldots ,N
 }
 \end{array}
  \eq
for the commuting Hamiltonians of the
Gaudin model. At the same time the quasiclassical limit of the spectral problem (\ref{q03}) provides
 some value $H^{\hbox{\tiny{CM}}}=E_0(\om)$ of the classical Calogero-Moser Hamiltonian
 \beq\label{q05}
 \begin{array}{c}
   \displaystyle{
 H^{\hbox{\tiny{CM}}}= \frac{1}{2}\sum\limits_{i=1}^N
p_i^2- \sum\limits_{i<j}^N \frac{g^2}{(q_i-q_j)^2}\,,\quad p_i={\dot q}_i
 }
 \end{array}
  \eq
 with the following identification of variables. The positions of classical particles $q_i$
 are identified with the marked points $z_i$ of the Gaudin model
 (as in the Schr\"odinger equation (\ref{q03})), the
 coupling constant $g$ is equal to
 the Planck constant $\hbar$, and the classical velocities ${\dot q}_i$ are
identified with the eigenvalues $H_i^{\hbox{\tiny{G}}}$
 of the quantum Gaudin Hamiltonians:
  \beq\label{q06}
    \displaystyle{
 q_j=z_j\,,\qquad g = \hbar \qquad \mbox{and} \qquad
\dot q_j=%\frac{1}{\hbar}\,
H_j^{\hbox{\tiny{G}}}\,, \ \
j=1\,,...\,,N\,.
 }
 \eq
 Similar fixation holds true for all higher Hamiltonians in involution
 of the Calogero-Moser model:
  \beq\label{q061}
    \displaystyle{
 H^{\hbox{\tiny{CM}}}_k=E_k(\om)\,,
 }
 \eq
   so that we finally obtain all action variables be fixed.
Equations (\ref{q061}) define some Lagrangian submanifold  in the classical
  $2N$-dimensional phase space.
  Its definition depends on the data of the initial KZ equations (\ref{q01}). For the first time the relation
  between the quantum ${\rm gl}_n$ Gaudin model and the classical Calogero-Moser system was mentioned in \cite{MTV} in the case
  without twists ($\om=0$).

  \paragraph{Lax matrix and Bethe ansatz.}
  In order to clarify the duality relation between the Gaudin model (\ref{q04}) and the
  Lagrangian submanifolds the classical Lax matrix of the
  Calogero-Moser system should be used. For the model (\ref{q05}) it is
  of the following form:
 \beq\label{q07}
 \begin{array}{c}
   \displaystyle{
L^{\hbox{\tiny{CM}}}_{ij}(\{\dot q_l\}\,,\{q_l\}\,, g)=
\delta_{ij}{\dot q}_i+g \frac{1-\delta_{ij}}{q_i-q_j}\,,\ \ \
i,j=1\,, \ldots \,,N\,.
 }
 \end{array}
  \eq
  \beq\label{q08}
 \begin{array}{c}
   \displaystyle{
M^{\hbox{\tiny{CM}}}_{ij}= \delta_{ij}\sum\limits_{k\neq
i}\frac{g}{(q_i-q_k)^2}-(1-\delta_{ij})
\frac{g}{(q_i-q_j)^2}\,,\ \ \ i,j=1\,, \ldots \,,N,
 }
 \end{array}
  \eq
  so that the $N\times N$ matrix Lax equation ${\dot L}=[L,M]$ is
  equivalent to the equations of motion
 \beq\label{q09}
 \begin{array}{c}
   \displaystyle{
\dot p_i=\ddot q_i =-\sum_{k\neq i}\frac{2g^2}{(q_i-q_k)^3}\,,
\quad i=1\,, \ldots \,,N\,.
 }
 \end{array}
  \eq
  Recall that the eigenvalues
  of the Lax matrix $\hbox{Spec}(L^{\hbox{\tiny{CM}}})=\{I_1, \ldots ,I_N\}$
   are the action variables for the model (\ref{q05}) since
  $\displaystyle{H^{\hbox{\tiny{CM}}}_k =\frac{1}{k}\, \tr
\left(L^{\hbox{\tiny{CM}}}\right)^k =\frac{1}{k}\sum_{i=1}^N I_i^k}$.

 To find the level of Hamiltonians $E_k(\omega)$ (\ref{q061})
 (or equivalently, the level of the action variables $I_k$), one can use the
 algebraic Bethe ansatz for the Gaudin model.
 The solution of the eigenvalue problems (\ref{q04}) is as follows:
 \beq\label{q10}
\begin{array}{c}
  \displaystyle{
%\frac{1}{\hbar}\,
H_i^{\hbox{\tiny{G}}}= \om+\sum\limits_{k\neq
i}^N\frac{\hbar}{z_i-z_k}+
\sum\limits_{\ga=1}^{M}\frac{\hbar}{\mu_\ga-z_i}\,,\quad
i=1, \ldots ,N\,,
 }
\end{array}
  \eq
 where the parameters $\{\mu_\al\,,\al=1, \ldots ,M\}$ are the Bethe roots
 satisfying the system of $M$ Bethe equations (BE)
 \beq\label{q11}
\begin{array}{c}
  \displaystyle{
2\om+ \hbar\sum\limits_{k=1}^N\frac{1}{\mu_\al-z_k}
=2\hbar\sum\limits_{\ga\neq\al}^{M}
\frac{1}{\mu_\al-\mu_\ga}\,,\quad \al=1, \ldots ,M\,.
 }
\end{array}
  \eq
The positive integer parameter $M$ is the number of overturned spins
in the Gaudin eigenvector $\psi\in{\mathcal H}$ (\ref{q04}).
In what follows we assume that $M\leq [N/2]$.
Using the identification (\ref{q06}) we can substitute
 ${\dot q}_j=H_i^{\hbox{\tiny{G}}}$ into the Lax matrix
(\ref{q07}) and compute the eigenvalues
$I_k$ (the action variables). It appears that on shell, i.e., when the Bethe
 equations (\ref{q11}) are satisfied, these eigenvalues take the form \cite{GZZ}
 \beq\label{q12}
 \begin{array}{c}
 \hbox{Spec} \, L^{\hbox{\tiny{CM}}}
 \left ( \{ %{\hbar}^{-1}
 H_j^{\hbox{\tiny{G}}} \},
 \left \{ z _j\right \} \,, \hbar \right )\Bigr |_{BE}
 =\big\{\underbrace{\om\,,\ldots\,,\om}_{N-M}\,, \,
 \underbrace{-\om\,,\ldots\,,-\om}_{M}\big\}\,.
\end{array}
  \eq
  That is the action variables of the classical model are twist parameters
  $\omega$, $-\omega$ with multiplicities given by
  the occupation numbers $N-M$, $M$ (the numbers of spins looking up and down in the state $\psi$).
   The identification of variables (\ref{q06}) can be viewed as initial conditions for
   the Calogero-Moser model (\ref{q07})--(\ref{q09}), i.e. the quantum-classical duality provides
some specific initial conditions for the classical model given by
intersection of two Lagrangian submanifolds. The first
   one is the $N$-dimensional level set of $N$ classical Hamiltonians in involution
   defined by (\ref{q12}) and the second one is the
   $N$-dimensional hyperplane $q_j=z_j$. (These are initial coordinates of the particles.)
In particular, if the twist is absent ($\omega =0$), then the first Lagrangian submanifold
is the zero set of the higher classical Hamiltonians.

  \paragraph{Determinant identities and factorization of the Lax matrix.}
The derivation of (\ref{q12}) is quite tricky (see \cite{GZZ}).
It uses some non-trivial determinant identities for the matrices of the form
   $L^{\hbox{\tiny{CM}}}
 \left ( \{ %{\hbar}^{-1}
 H_j^{\hbox{\tiny{G}}}(z_k,\mu_k,\hbar,\om) \},
 \left \{ z _j\right \} \,, \hbar \right )$.
 For the example discussed above the identity looks as follows:
  \beq\label{q13}
  \displaystyle{
 \det_{N\times N}
 \Bigl ({\mathcal{L}}-\la I\Bigr )=(\om-\la)^{N-M}
\det_{M\times M} \Bigl ({\widetilde {\mathcal{L}}} -\la I\Bigr )\,,
 }
  \eq
 where $I$ is the identity matrix and
 \beq\label{q14}
\begin{array}{c}
 \displaystyle{
  {\mathcal{L}}_{ij}=
\delta_{ij}\left(\om+\sum\limits^N_{k\neq
i}\frac{\hbar}{q_i-q_k}+\sum\limits_{\ga=1}^M\frac{\hbar}{\mu_\ga-q_i}
\right)+(1-\delta_{ij})\frac{\hbar}{q_i-q_j}\,,\quad i,j=1, \ldots ,N }
\end{array}
  \eq
 \beq\label{q15}
\begin{array}{c}
\displaystyle{{\widetilde {\mathcal{L}}}_{\al\be} =
\delta_{\al\be}\left(
\om-\sum\limits^M_{\ga\neq\al}\frac{\hbar}{\mu_\al\!-\!\mu_\ga}
-\sum\limits^N_{k=1}\frac{\hbar}{q_k\!-\!\mu_\al}\right)+
\left(1-\delta_{\al\be}\right) \frac{\hbar}{\mu_\al\!-\!\mu_\be}\,,
 \quad \al,\be=1, \ldots ,M\,.
}
\end{array}
  \eq
Identities of such type
appear in studies of scalar products of Bethe vectors in quantum integrable models
\cite{Slavnov,Claeys}.
 The proof of these identities is based on another non-trivial phenomenon --
factorization of the Lax matrices \cite{Hasegawa,VZ}. For example, consider
  the matrix $\mL$ (\ref{q14}) for $M=0$. It can be represented in the
  following form\footnote{In this paper the notations for $V,C_0$ differ
  from those in \cite{GZZ} by transposition.}:
 \beq\label{q16}
 \begin{array}{c}
 \displaystyle{
  \mL=\om I+\hbar (D^0)^{-1}VC_0V^{-1}D^0=(D^0)^{-1}V
  \Big(\om I+\hbar C_0\Big)V^{-1}D^0\,,\qquad V_{ij}(q)=q_i^{j-1}\,,
}
\end{array}
  \eq
 \beq\label{q17}
 \begin{array}{c}
 \displaystyle{
  % V_{ij}(q)=q_i^{j-1}\,,  \quad
    (D^0)_{ij}=\delta_{ij}\prod\limits_{k\neq i}^N(q_i-q_k)\,,
  \quad (C_0)_{ij}=i\,\delta_{i+1,j}\,,
%\quad (\ti C)_{ij}=\frac{1+(-1)^j}{2}\, \delta_{i+1,j}\,,
}
\end{array}
  \eq
  where
  %${\ }^T$ means matrix transposition,
  $V$ is the Vandermonde matrix,
   $D^0$ is diagonal and $C_0$ is the upper-triangular matrix.
In factorization formulas for Lax matrices for Calogero-Moser models associated with the
root systems of types $B,C,D$ the upper-triangular matrix
\beq\label{q17a}
\ti C_{ij}=\frac{1+(-1)^j}{2}\, \delta_{i+1,j}
\eq
is also used, see (\ref{w11}), (\ref{w15}).
   From (\ref{q16}) it immediately follows that such matrix $\mL$ has
   all eigenvalues equal to $\om$. It is the statement (\ref{q13}) for $M=0$.
   Below we use some modifications of the above determinant identities and factorization formulae.

   \paragraph{From duality to correspondence: supersymmetric generalization.}
    In the supersymmetric case the main statement of the duality relation (\ref{q12}) is that it
   holds  true for ${\rm gl}(1|1)$ and ${\rm gl}(0|2)$ supersymmetric Gaudin models \cite{Kulish,Essler, Gonz,Arn,Kazakov-Zabrodin,BeR} as well as for
${\rm gl}(2|0)$. In the general case one should take into consideration all
   $m+n+1$ models associated with the superalgebras ${\rm gl}(m|n)$
   with $m+n$ fixed \cite{TsuboiZZ}.
  In this respect a single classical system corresponds to a number
  of quantum models.

  Notice that duality (\ref{q12}) could be used (in principle)
  for a direct solution of the quantum spectral problem (\ref{q04})
  without using the Bethe ansatz equations (\ref{q11}). Indeed, one may write down
  a system of algebraic equations for the Lax matrix (\ref{q07}) to have
  the eigenvalues (\ref{q12}). Solving this systems with respect to velocities one
  finds the spectrum (\ref{q04}) as $H_j^{\hbox{\tiny{G}}}=\dot q_j(\{q_k\},\om,\hbar)$. However, it follows from the above statement
  that a given solution for velocities may correspond to one or another Gaudin model
  among the $m+n+1$
  models, and it is not clear in general to which one. To clarify the underlying
  combinatorics is an interesting open problem. At quantum level the relation between (q)KZ equations related
  to supersymmetric Lie algebras and the
  Calogero-Ruijsenaars many-body systems was studied in \cite{GrZZ} for $A$-type models.

   \paragraph{Calogero-Moser systems and Gaudin models with boundary.}
  In our previous paper \cite{VZZ} we studied the duality for Calogero-Moser models associated with the
  classical root systems of simple Lie algebras, i.e. to the
  root systems of $B_N$, $C_N$ and $D_N$ types.
  These models were introduced in \cite{OP}. In the rational
  case the Hamiltonian is of the following form:
    \begin{equation}
 \label{w01}
 H = \frac{1}{2} \sum_{a=1}^{N} p_a^2 - g_2^2
\sum_{a<b}^{N} \Big(\frac1{(q_a-q_b)^2} + \frac1{(q_a+q_b)^2}\Big) -
g_4^2\sum_{a=1}^{N} \frac1{(2q_a)^2} - g_1^2 \sum_{a=1}^{N}
\frac1{q_a^2}.
 \end{equation}
 It depends on three free parameters,
 the coupling constants $g_2$, $g_4$ and $g_1$. Two of them ($g_4$ and $g_1$) can be unified in
 (\ref{w01}) into a single combination $g_4^2/4+g_1^2$. We do not do that since the two last terms
 are associated with roots of different types, and
 in the trigonometric (and elliptic) extensions these terms are
 different. The root systems  $B_N$, $C_N$ and $D_N$ are distinguished
 by values of the coupling constants, see (\ref{w06}).

 The size of the matrices participating in the
Lax pair for the model (\ref{w01}) is $2N\times 2N$ ($C,D$) or $(2N+1)\times(2N+1)$ ($B$). The corresponding factorization formulas of the
type (\ref{q16})--(\ref{q17}) were derived in \cite{VZ}, see (\ref{w11}), (\ref{w15}).
 Based on this knowledge we showed in \cite{VZZ} that  (\ref{w01}) is quantum-classically dual to
 the boundary Gaudin magnet  -- the Gaudin limit of the $XXX$ quantum spin chain with (some special) boundary conditions introduced by Sklyanin \cite{Skl}. Using
 the algebraic Bethe ansatz for the boundary Gaudin models \cite{Hikami,Luk} we derived
 the underlying matrix identities (\ref{w53}) and (\ref{w50}) of type
 (\ref{q13}) and computed the levels of the classical Hamiltonians. In contrast to $A_{N-1}$ type Calogero-Moser model (\ref{q05}),
  all eigenvalues of $BCD$-type Lax matrix are equal to zero:
  %since there is no twist matrices in the related spin chain (and Gaudin model):
  %
 \beq\label{q20}
 \begin{array}{c}
 \det
 \Big( L^{\hbox{\tiny{CM}}}
 \left ( \{ %{\hbar}^{-1}
 H_j^{\hbox{\tiny{G}}} \},
 \left \{ z _j\right \} \,, \hbar \right )\Bigr |_{BE}-\lambda I
 \Big)=(-\lambda)^{r}\,,
\end{array}
  \eq
 where $r=2N$ for $C_N$, $D_N$ root systems and  $r=2N+1$ for $B_N$.

   \paragraph{Purpose of the paper} \hspace{-3mm}
   is to extend the quantum-classical duality  between the
   quantum  boundary Gaudin magnet and the
   classical  Calogero-Moser model of types $BCD$ to the correspondence.
Namely, a single classical model (associated to the root system of type $B$, $C$ or $D$)
   is related to three supersymmetric Gaudin magnets associated with
   superalgebras ${\rm gl}(2|\,0)$,  ${\rm gl}(1|\,1)$
   and ${\rm gl}(0|\,2)$. The eigenvalues and Bethe equations are described in the next section.
It should be stressed that
our previous construction for ${\rm gl}(2|\,0)$ \cite{VZZ} can not be
straightforwardly extended
 to the supersymmetric case.
Some additional computational tricks
%and properties of the Frobenius companion matrix
are required for the proof.
Some details of the proof are
  given in Section 3. The details of the Lax representation for the
  Calogero-Moser models of types $BCD$,
 including factorization formulae and determinant identities,
   can be found in the appendix. In the Conclusion we summarize the obtained results.

%%%%%%%%%%%%%%%%%%%%%%%%%%%%%%%%%%%%%%%%%%%%%%%%%%%%%%%%%%%%%%%%%%%%%%%%%%%%%%%%%%%%%%%
%%%%%%%%%%%%%%%%%%%%%%%%%%%%%%%%%%%%%%%%%%%%%%%%%%%%%%%%%%%%%%%%%%%%%%%%%%%%%%%%%%%%%%%

\section{Supersymmetric Gaudin model with boundary}
\setcounter{equation}{0}

We consider ${\rm gl}(m|n)$ Gaudin magnets with open boundary conditions and $m+n=2$, which arise
 by applying Gaudin limit
to the $\mathbb{Z}_2$-graded quantum
spin chains. The rational models are defined through the graded permutation operator
 \beq\label{q30}
 \displaystyle{
 {\bf P} = \sum\limits_{i,j=1}^{m+n} (-1)^{p(j)} E_{ij}\otimes E_{ji}\,,\qquad
 {\bf P}(x \otimes y) =(-1)^{p(x)p(y)} (y \otimes x)\,,\ x,y\in\mC^{m|n}\,.
 }
\eq
 See the notation in (\ref{w17})--(\ref{w24}). The supersymmetric
 Yang's $R$-matrix is of the form
\beq \label{w18}
 \displaystyle{
R(u) = {\bf I}+\frac{\eta}{u}\,{\bf P}\,.
 }
\eq
 It satisfies the graded Yang-Baxter equation
\beq \label{w19}
 \displaystyle{
R_{12}(u_1-u_2)R_{13}(u_1-u_3)R_{23}(u_2-u_3)=R_{23}(u_2-u_3)R_{13}(u_1-u_3)R_{12}(u_1-u_2).
}
\eq
For the construction of integrable spin chains with open boundary conditions one needs
%the exchange relations
%
%\beq \label{w20}
%\displaystyle{
%R_{12}(u_1-u_2)T_1(u_1)T_2(u_2)=T_2(u_2)T_1(u_1)R_{12}(u_1-u_2)
%}
%\eq
%
the reflection equations for boundary $K$-matrices $K^{\pm}(u)$:
\beq \label{w21}
\begin{array}{c}
\displaystyle{
R_{12}(u_1-u_2)K_1^{-}(u_1)R_{21}(u_1+u_2)K_2^{-}(u_2) =
K_2^{-}(u_2)R_{12}(u_1+u_2)K_1^{-}(u_1)R_{21}(u_1-u_2)\,,
}
\end{array}
\eq
$$
\begin{array}{c}
\displaystyle{
R_{12}(u_2-u_1)(K_1^{+}(u_1))^{t_1}R_{21}(-u_1-u_2-(m-n)\eta)(K_2^{+}(u_2))^{t_2} =
}
\\ \ \\
\displaystyle{
(K_2^{+}(u_2))^{t_2}R_{12}(-u_1-u_2-(m-n)\eta)(K_1^{+}(u_1))^{t_1}R_{21}(u_2-u_1)\,,
}
\end{array}
 $$
where $t_i$ means super transposition in the $i$-th tensor component  (\ref{w23}).

In this paper we consider the following diagonal
solutions\footnote{General $K$-matrices presumably correspond to the general Lax matrix
of the Calogero-Moser model \cite{Feher}, which has no restriction (\ref{w05}) for the coupling constants.
 However the factorization properties and determinant identities
 are yet unknown for the general Lax pair. } of the
reflection equations (\ref{w21}):
\beq \label{w25}
\begin{array}{c}
K^-(u) = \left(\begin{array}{cc}
\displaystyle{1+\frac{\al \eta}{u} }&0\\ \ \\
0& \displaystyle{-1+\frac{\al \eta }{u} }
\end{array}
\right),\quad
K^+(u)=\left(\begin{array}{cc}
\displaystyle{ 1-\frac{\be \eta }{u+\frac{m-n}{2}\eta} }&0\\ \ \\
0&\displaystyle{-1-\frac{\be \eta}{u+\frac{m-n}{2}\eta} }
\end{array}
\right).
\end{array}
\eq
Relations (\ref{w19}) (\ref{w21}) guarantee that the transfer matrices
\beq \label{w26}
\displaystyle{
{\bf{T}}(u) = \hbox{str}_0 \Big(K^{+}_0(u)R_{01}(u-z_1)\ldots R_{0N}
(u-z_N)K^{-}_0(u)R_{0N}(u+z_N)\ldots R_{01}(u+z_1)\Big)
 }
\eq
commute for different values of the spectral parameter: $[{\bf{T}}(u),{\bf{T}}(v)]=0$ for all $u,v$.
The Gaudin model appears in the limit $\eps \rightarrow 0$
after the substitution $\eta=\eps \hbar$:
\beq \label{w27}
{\bf{T}}(u)= (-1)^{p(1)}+(-1)^{p(2)} +
\eps \hbar \ga(u) + \eps^2 \hbar^2 {\bf{T}}^{ \rm G}(u) + O(\eps^3),
\eq
where $\ga(u)$ is a scalar function. The notation $p(i)$ is defined in (\ref{w17}).
The expression
\beq \label{w28}
  \displaystyle{
  {{\bf T}}^{\hbox{\tiny{G}}}(u) =-\frac{2\alpha \beta}{u^2} + \frac{1}{\hbar}
 \sum\limits_{i=1}^{N}\Big(\frac{ {\bf H}_i^{\hbox{\tiny{G}}} }{u-z_i}-
 \frac{ { {\bf H}_i^{\hbox{\tiny{G}}} }}{u+z_i}\Big)
 }
 \eq
is the Gaudin transfer matrix, and the operators ${\bf H}_i^{\hbox{\tiny{G}}}$ are commuting Gaudin Hamiltonians:
\beq \label{w281}
 \displaystyle{
 \frac{1}{\hbar}\,{\bf H}_i^{\hbox{\tiny{G}}} =\frac{(2\xi+(-1)^{p(1)}-(-1)^{p(2)}) \sigma_3^{(i)} }{2z_i}+
  \sum\limits_{k \neq i}^{N}\Big(\frac{{\bf P}_{ik}}{z_i-z_k}+\frac{\sigma_3^{(i)}
{\bf P}_{ik}\sigma_3^{(i)} }{z_i+z_k}\Big)
 \,,\qquad \xi = \alpha -\beta\,,
 }
 \eq
 where ${\bf P}_{ik}$ are the graded permutation operators (\ref{q30})
 exchanging  $i$-th and $k$-th tensor components
 of the Hilbert space $(\mC^{m|n})^{\otimes N}$.
 The Gaudin spectral problems are
\beq \label{w282}
 \displaystyle{
 {\bf H}_i^{\hbox{\tiny{G}}}\psi = H_i^{\hbox{\tiny{G}}}\psi .
 }
 \eq
The eigenvalues $H_{i}^{\rm G}$ of these operators
are obtained via the algebraic Bethe ansatz technique \cite{Essler}.

 The Gaudin model  (\ref{w28}), (\ref{w281}) with $2N$ marked points
 $z_1,\ldots ,z_N,-z_1,\ldots ,-z_N$ will
 be shown to be dual to the Calogero-Moser models of types $C_N$ and $D_N$. Besides this case, we
 also need the one related to the $B_N$ root system.
 It comes from the transfer matrix (\ref{w28}) by substituting
 $N\rightarrow N+1$ and $z_{N+1}=0$. Then we have $2N+1$ marked points
 $z_1,\ldots ,z_N,0, -z_1,\ldots ,-z_N$.
 In this case we fix the parameter  $\xi=p(1)-p(2)$. Consider the solutions of the
 eigenvalue problem (\ref{w282}) in all the cases \cite{Essler}.

 \noindent\underline{\bf $2N$ marked points.}
 The eigenvalues of the Gaudin Hamiltonians (\ref{w281}) are of the form
\beq \label{w29}
\begin{array}{c}
\displaystyle{
 \frac{1}{\hbar}\, H_i^{\hbox{\tiny{G}}} = \frac{\xi-p(1)+p(2)}{z_i}
 +\sum\limits_{k\neq i}^{N}\left(\frac{(-1)^{p(1)}}{z_i-z_k}+\frac{(-1)^{p(1)}}{z_i+z_k} \right)
-\sum\limits_{l=1}^{M}\left(\frac{(-1)^{p(1)}}{z_i-\mu_l}+\frac{(-1)^{p(1)}}{z_i+\mu_l} \right),
}
\end{array}
\eq
where $\xi=\al-\be$. The set of Bethe roots $\{\mu\}_{M}=\{\mu_1,\ldots ,\mu_M\}$
 satisfy the system of $M$ Bethe equations ($l=1,\ldots ,M$)
\beq \label{w30}
\begin{array}{c}
\displaystyle{
\frac{2\xi}{\mu_l} + (-1)^{p(1)}\sum\limits_{k=1}^{N}
\left(\frac{1}{\mu_l-z_k}+\frac{1}{\mu_l+z_k}\right) =
}
\\ \ \\
\displaystyle{
= \Big((-1)^{p(1)} + (-1)^{p(2)}\Big)\left(\frac{1}{\mu_l} +
\sum\limits_{k \neq l}^{M}\left(\frac{1}{\mu_l-\mu_k}+\frac{1}{\mu_l+\mu_k} \right) \right)\,.
}
\end{array}
\eq
Let us write down (\ref{w29}) and (\ref{w30}) explicitly for different superalgebras:

\underline{${\rm gl}(2|0)$ case:}
\beq \label{w31}
\begin{array}{c}
\displaystyle{
\frac{1}{\hbar}\,H_{i}^{\rm G(2|0)}(\{z\}_N,\{\mu\}_M,\xi) = \frac{\xi}{z_i} +\sum\limits_{k\neq i}^{N}\left(\frac{1}{z_i-z_k}+\frac{1}{z_i+z_k} \right)
-\sum\limits_{l=1}^{M}\left(\frac{1}{z_i-\mu_l}+\frac{1}{z_i+\mu_l} \right),
}
\end{array}
\eq
\beq \label{w32}
\begin{array}{c}
\displaystyle{
2\frac{\xi}{\mu_l} + \sum\limits_{k=1}^{N}\left(\frac{1}{\mu_l-z_k}+\frac{1}{\mu_l+z_k}\right)
= 2\left(\frac{1}{\mu_l} +\sum\limits_{k \neq l}^{M}\left(\frac{1}{\mu_l-\mu_k}+\frac{1}{\mu_l+\mu_k} \right) \right).
}
\end{array}
\eq

\underline{${\rm gl}(1|1)$ case:}
\beq \label{w33}
\begin{array}{c}
\displaystyle{
\frac{1}{\hbar}\,H_{i}^{\rm G(1|1)}(\{z\}_N,\{\mu\}_M,\xi)
}
\\ \ \\
\displaystyle{
=\frac{\xi+1}{z_i} +\sum\limits_{k\neq i}^{N}\left(\frac{1}{z_i-z_k}+\frac{1}{z_i+z_k} \right)
-\sum\limits_{l=1}^{M}\left(\frac{1}{z_i-\mu_l}+\frac{1}{z_i+\mu_l} \right),
}
\end{array}
\eq
\beq \label{w34}
\begin{array}{c}
\displaystyle{
2\frac{\xi}{\mu_l} + \sum\limits_{k=1}^{N}\left(\frac{1}{\mu_l-z_k}+\frac{1}{\mu_l+z_k}\right) =0\,.
}
\end{array}
\eq

\underline{${\rm gl}(0|2)$ case:}
\beq \label{w35}
\begin{array}{c}
\displaystyle{
\frac{1}{\hbar}\,H_{i}^{\rm G(0|2)}(\{z\}_N,\{\mu\}_M,\xi) = \frac{\xi}{z_i} -\sum\limits_{k\neq i}^{N}\left(\frac{1}{z_i-z_k}+\frac{1}{z_i+z_k} \right)
+\sum\limits_{l=1}^{M}\left(\frac{1}{z_i-\mu_l}+\frac{1}{z_i+\mu_l} \right),
}
\end{array}
\eq
\beq \label{w36}
\begin{array}{c}
\displaystyle{
-2\frac{\xi}{\mu_l} + \sum\limits_{k=1}^{N}\left(\frac{1}{\mu_l-z_k}+\frac{1}{\mu_l+z_k}\right)
= 2\left(\frac{1}{\mu_l} +\sum\limits_{k \neq l}^{M}\left(\frac{1}{\mu_l-\mu_k}+\frac{1}{\mu_l+\mu_k} \right) \right).
}
\end{array}
\eq
In all these cases $H_{i}^{\rm G(m|n)}=H_{i}^{\rm G(m|n)}(\{z\}_N,\{\mu\}_M,\xi)$, i.e. the eigenvalues depend on $N$ marked points $z_i$, $M$ Bethe roots $\mu_j$ and the parameter $\xi$.

 \noindent\underline{\bf $2N+1$ marked points.}
 Here we consider the Gaudin model (\ref{w27})--(\ref{w28})
 with $N+1$ spins and $\xi = p(2)-p(1)$, $z_{N+1}=0$. Then
(\ref{w29})--(\ref{w30}) acquire the form
\beq \label{w037}
\begin{array}{c}
\displaystyle{
\frac{1}{\hbar}\,\widetilde{H}^{\rm{G}}_i(\{z\}_N,\{\mu\}_M)
}
\\ \ \\
\displaystyle{
 =(-1)^{p(1)}\left(
\frac{2}{z_i}+ \sum\limits_{k\neq i}^{N}\left(\frac{1}{z_i-z_k}+\frac{1}{z_i+z_k}\right) -
\sum\limits_{l=1}^{M}\left( \frac{1}{z_i-\mu_l}+\frac{1}{z_i+\mu_l}\right)\right)\,,
}
\end{array}
\eq
\beq \label{w038}
\begin{array}{c}
\displaystyle{
(-1)^{p(1)}\sum\limits_{k=1}^{N}\left(\frac{1}{\mu_l-q_k}+\frac{1}{\mu_l+q_k}
\right) = \left((-1)^{p(1)}+(-1)^{p(2)}\right) \sum\limits_{k \neq l}^{M}
\left( \frac{1}{\mu_l-\mu_k}+\frac{1}{\mu_l+\mu_k}\right)\,.
}
\end{array}
\eq
For each superalgebra (\ref{w037})--(\ref{w038}) we have:

\underline{${\rm gl}(2|0)$ case:}
\beq \label{w039}
\begin{array}{c}
\displaystyle{
\frac{1}{\hbar}\,\widetilde{H}^{\rm{G}(2|0)}_i(\{z\}_N,\{\mu\}_M) =\frac{2}{z_i} +
\sum\limits_{k\neq i}^{N}\left(\frac{1}{z_i-z_k}+\frac{1}{z_i+z_k}\right) -
\sum\limits_{l=1}^{M}\left( \frac{1}{z_i-\mu_l}+\frac{1}{z_i+\mu_l}\right),
}
\end{array}
\eq
\beq \label{w040}
\begin{array}{c}
\displaystyle{
\sum\limits_{k=1}^{N}\left(\frac{1}{\mu_l-q_k}+\frac{1}{\mu_l+q_k}\right) =
2 \sum\limits_{k \neq l}^{M}\left( \frac{1}{\mu_l-\mu_k}+\frac{1}{\mu_l+\mu_k}\right).
}
\end{array}
\eq

\underline{${\rm gl}(1|1)$ case:}
\beq \label{w041}
\begin{array}{c}
\displaystyle{
\frac{1}{\hbar}\,\widetilde{H}^{\rm{G}(1|1)}_i(\{z\}_N,\{\mu\}_M) =\frac{2}{z_i} +
\sum\limits_{k\neq i}^{N}\left(\frac{1}{z_i-z_k}+\frac{1}{z_i+z_k}\right) -
\sum\limits_{l=1}^{M}\left( \frac{1}{z_i-\mu_l}+\frac{1}{z_i+\mu_l}\right),
}
\end{array}
\eq
\beq \label{w042}
\begin{array}{c}
\displaystyle{
\sum\limits_{k=1}^{N}\left(\frac{1}{\mu_l-q_k}+\frac{1}{\mu_l+q_k}\right) = 0.
}
\end{array}
\eq

\underline{${\rm gl}(0|2)$ case:}
\beq \label{w043}
\begin{array}{c}
\displaystyle{
\frac{1}{\hbar}\,\widetilde{H}^{\rm{G}(0|2)}_i(\{z\}_N,\{\mu\}_M) =
-\frac{2}{z_i} - \sum\limits_{k\neq i}^{N}\left(\frac{1}{z_i-z_k}+\frac{1}{z_i+z_k}\right) +\sum\limits_{l=1}^{M}\left( \frac{1}{z_i-\mu_l}+\frac{1}{z_i+\mu_l}\right),
}
\end{array}
\eq
\beq \label{w044}
\begin{array}{c}
\displaystyle{
\sum\limits_{k=1}^{N}\left(\frac{1}{\mu_l-q_k}+\frac{1}{\mu_l+q_k}\right)
=  2\sum\limits_{k \neq l}^{M}\left( \frac{1}{\mu_l-\mu_k}+\frac{1}{\mu_l+\mu_k}\right).
}
\end{array}
\eq

%%%%%%%%%%%%%%%%%%%%%%%%%%%%%%%%%%%%%%%%%%%%%%%%%%%%%%%%%%%%%%%%%%%%%%%%%%%%%%%%%%%%%%%%%%%%%%%%

\section{Proof of the correspondence}
\setcounter{equation}{0}

The statement of the correspondence is that relation (\ref{q20}) holds true for the Lax matrices
 (\ref{w02}) of $BCD$ types, where velocities of the Calogero-Moser particles are identified with
 the eigenvalues of ${\rm gl}(2|0)$ or ${\rm gl}(1|1)$ or ${\rm gl}(0|2)$ Gaudin model Hamiltonians given in
 (\ref{w29}) or (\ref{w037}) for $C,D$ and $B$ root systems respectively. More precisely,
 make the following identifications:
 \beq\label{a027}
    \displaystyle{
 z_j=q_j\,,
\quad j=1,\ldots,N\,,
 }
 \eq
 \beq\label{a028}
    \displaystyle{
\dot q_j= H_j^{{\rm G}(m|n)}(\{q\}_N,\{\mu\}_M,\xi)
\quad\hbox{or}\quad \dot q_j= {\widetilde H}_j^{{\rm G}(m|n)}(\{q\}_N,\{\mu\}_M)\,.
\quad j=1,\ldots,N\,.
 }
 \eq
  Next, consider the Lax matrix $$L(\{\dot q_j \}_N, \{q_j \}_N|\,g_1,g_2,g_4)$$ from
  (\ref{w02})--(\ref{w06}).
  %The coupling constants are chosen as given in (\ref{w101}) or (\ref{w141}),
  The size of the Lax matrix is equal to
  $r=2N$ for $C_N$, $D_N$ root systems and  $r=2N+1$ for $B_N$.
  Then we are going to prove the following statement:
 \beq\label{a029}
 \begin{array}{c}
 \det\limits_{2N\times 2N}
 \Big( L
 \left ( \left\{ %{\hbar}^{-1}
 H_j^{{\rm G}(m|n)}(\{q\}_N,\{\mu\}_M,\xi) \right\}_N,
 \left \{ q _j\right \}_N \Big|\,g_1,g_2,g_4) \right )\Bigr |_{BE\,(\ref{w30})}-\lambda I
 \Big)=\lambda^{2N}\,,
\end{array}
  \eq
where
 \beq\label{a0291}
 \begin{array}{l}
 \hbox{for}\quad C_N:\ \ g_1=0\,,g_2=\hbar\,,g_4=\sqrt{2}\hbar(\xi-p(1)+p(2))\,,
 \\ \ \\
 \hbox{for}\quad D_N:\ \ \xi=p(1)-p(2)\,, g_1=0\,,g_2=\hbar\,,g_4=0
\end{array}
  \eq
 and
 \beq\label{a0292}
 \begin{array}{c}
 \det\limits_{(2N\!+\!1)\times(2N\!+\!1)}\!
 \Big( L
 \left ( \left\{ %{\hbar}^{-1}
 {\widetilde H}_j^{{\rm G}(m|n)}(\{q\}_N,\{\mu\}_M) \right\}_N,
 \left \{ q _j\right \}_N \Big|\,\sqrt{2}\hbar,\hbar,0 \right )\Bigr |_{BE\,(\ref{w038})}
 \!-\lambda I
 \Big)=-\lambda^{2N\!+\!1}\,.
\end{array}
  \eq
The proof for ${\rm gl}(2|0)$ with parameters (\ref{w101}) or (\ref{w141})
was given in \cite{VZZ}. Here we prove the cases
of the superalgebras ${\rm gl}(1|1)$ and ${\rm gl}(0|2)$. We will see
that the latter can be reduced to ${\rm gl}(2|0)$ using special properties of the Lax matrices.
The proof for ${\rm gl}(1|1)$ requires
additional computational trick based on the usage of the Frobenius matrix and matching of the
parameters.

%%%%%%%%%%%%%%%%%%%%%%%%%%%%%%%%%%%%%%%%%%%%%%%%%%%%%%%%%%%%%%%%%%%%%%%%%%%%%%%%%%%%%%%%%%%%%%%%

\subsection{$C_N$ and $D_N$ root systems for ${\rm gl}(1|1)$ superalgebra}

Here $m=n=1$. We begin with the $C_N$ root system since $D_N$ comes as a particular case of it.
 It follows from (\ref{a0291}) and (\ref{w17}) that $p(1)=0$, $p(2)=1$ and
 $g_1=0$, $g_2=\hbar$, $g_4=\sqrt{2}\hbar \, (\xi+1)$.

 Introduce the short-hand notation for the Lax matrix
 of size $2N\times 2N$ entering (\ref{a029}):
 \beq \label{aa2}
\begin{array}{c}
\displaystyle{
\mathcal{L}^{1|1} = L\left(\{H_j^{\rm{G}(1|1)}(\{q\}_N,\{\mu\}_M,\xi)\},
\{q\}_N |\, 0,\hbar,\sqrt{2}\hbar \left(\xi+1\right)\right)\,.
}
\end{array}
\eq
 By comparing  (\ref{w31}) and (\ref{w33}) we conclude that
 \beq \label{aa3}
\begin{array}{c}
\displaystyle{
H_j^{\rm{G}(1|1)}(\{q\}_N,\{\mu\}_M,\xi)=H_j^{\rm{G}(2|0)}(\{q\}_N,\{\mu\}_M,\xi+1)\,.
}
\end{array}
\eq
Therefore,
 \beq \label{aa4}
\begin{array}{c}
\displaystyle{
\mathcal{L}^{1|1} = L\left(\{H_j^{\rm{G}(2|0)}(\{q\}_N,
\{\mu\}_M,\xi+1)\},\{q\}_N |\, 0,\hbar,\sqrt{2}\hbar \left(\xi+1\right)\right)\,,
}
\end{array}
\eq
 that is the Lax matrix has the form (\ref{w51}) written
 for ${\rm gl}(2|\,0)$ case but with $\xi$ replaced by $\xi+1$. However, we can not use the duality statement
 for ${\rm gl}(2|\,0)$ case here since the Bethe equations in ${\rm gl}(1|1)$ case
  (\ref{w34}) differ from those (\ref{w32}) for ${\rm gl}(2|\,0)$.
  We can use this argument when $M=0$ only (then the set of Bethe equations is empty). For $M=0$
  we get $\mathcal{L}^{1|1}=L'(\xi\rightarrow\xi+1)$ for $L'$ (\ref{w11}). The duality relation (\ref{a029}) then
  follows immediately from the explicit form of matrices $C_0$ and
  $\tilde C$ (\ref{q17a}) because they are
  upper-triangular.

  Suppose $M\geq 1$. Let us apply the determinant identity (\ref{w53}) to (\ref{aa4}). We get:
 \beq\label{a3}
 \displaystyle{
 \det\limits_{2N \times 2N}\left(\mathcal{L}^{1|1}-\la I
 \right) = \la^{2N-2M}\det\limits_{2M\times 2M}\left(\tilde{\mathcal{L}}^{1|1}-\la I\right)\,,
 }
 \eq
 where $2M\times 2M$ dual matrix is of the form
 \beq \label{aa5}
\begin{array}{c}
\displaystyle{
\ti{\mathcal
 L}^{1|1}=L\Big(\{H_j^{\rm{G}(2|0)}(\{\mu\}_M,\{q\}_N,-
 \xi )\},\{\mu \}_M|\,0,\hbar,-\sqrt{2}\, \hbar \, \xi \Big)\,.
}
\end{array}
\eq
 Next, let us impose the Bethe equations (\ref{w34}) in (\ref{aa5}) (take it ``on-shell'').
 This yields
\beq \label{a4}
\displaystyle{
\tilde{\mathcal{L}}^{1|1}\Big|_{BE (\ref{w34})} =
 L\left( \{H_j^{\rm{G}(2|0)}\left(\{\mu\}_M,\{\emptyset\},\xi\right),\{\mu\}_M |\,0,\hbar,-\sqrt{2}\hbar \xi \right)\,.
}
\eq
 Using explicit form of (\ref{w31}), we have:
\beq \label{a41}
\displaystyle{
H_j^{\rm{G}(2|0)}\left(\{\mu\}_M,\{ q\}_N,\xi\right)=
H_j^{\rm{G}(2|0)}\left(\{\mu\}_M,\{ q\}_N,-\xi\right)+\frac{2\xi}{\mu_j}\,,
}
\eq
so that
\beq \label{a42}
\displaystyle{
\tilde{\mathcal{L}}^{1|1}_{ij}\Big|_{BE (\ref{w34})} =
 L_{ij}\left( \{H_k^{\rm{G}(2|0)}\left(\{\mu\}_M,\{\emptyset\},
 -\xi\right),\{\mu\}_M |\,0,\hbar,-\sqrt{2}\hbar \xi \right)
 \pm\delta_{ij}\frac{2\xi}{\mu_j}
}
\eq
(for $i,j=1,\ldots ,2M$),
where the sign $+$ is for $1\leq i\leq M$ and the sign $-$ for $M+1\leq i\leq 2M$.
The first term is transformed using the factorization formula (\ref{w11}):
\beq \label{a43}
\begin{array}{c}
\displaystyle{
 L\left( \{H_j^{\rm{G}(2|0)}\left(\{\mu\}_M,\{\emptyset\},
 -\xi\right),\{\mu\}_M |\,0,\hbar,-\sqrt{2}\hbar \xi \right)
}
\\ \ \\
\displaystyle{
=\hbar (D^{0})^{-1}V\left(C_0-(1+2\xi)\tilde{C} \right)V^{-1}D^{0}\,.
}
\end{array}
\eq
The second term in (\ref{a42}) is a diagonal matrix, which can represented
through the Frobenius companion matrix
(\ref{w162}) as follows:
\beq \label{a44}
\begin{array}{c}
\displaystyle{
2\xi\hbar\, (D^{0})^{-1}V\left( J^{-1}\right)V^{-1}D^{0}\,,
}
\end{array}
\eq
 where the set of variables $\{x_k\}$ defining $J$ is $(\mu_1,...,\mu_M,-\mu_1,...,-\mu_M)$.
 Finally,
we obtain
\beq \label{a5}
\begin{array}{c}
\displaystyle{
\tilde{\mathcal{L}}^{1|1}
\Big|_{BE (\ref{w34})} = \hbar (D^{0})^{-1}V\left(C_0-(1+2\xi)\tilde{C} +2\xi J^{-1}\right)V^{-1}D^{0},
}
\end{array}
\eq
where the
matrices $C_0$ and $\tilde{C}$ are as in (\ref{q17}), (\ref{q17a}) (their size
is $2M\times 2M$), and the matrix $J^{-1}$
(\ref{w16}) is of the form
\beq \label{a6}
J^{-1} =\left( \begin{array}{ccccccc}
0 & 1 & 0& . & .& . &0\\
-e_{2M-2}(\mu_1^{-1},..,\mu_M^{-1},-\mu_1^{-1},..,-\mu_M^{-1}) & 0 & 1& .& .& . & 0\\
0 & 0 & 0& .&. & . & 0\\
.&.&.&& &.&0\\
.&.& & .& &.&0\\
.&.& & & .&.&1\\
-e_0(\mu_1^{-1},..,\mu_M^{-1},-\mu_1^{-1},..,-\mu_M^{-1})&0&0&.&.&.&0
\end{array} \right).
\eq
Here $e_k$ are elementary symmetric polynomials of the indicated variables.
Then we obtain:
\beq \label{a7}
\begin{array}{c}
\displaystyle{
 \det\limits_{2M\times 2M}\left(\tilde{\mathcal{L}}^{1|1}\Big|_{BE (\ref{w34})}-\la I\right) =
 \det\limits_{2M\times 2M}\left(C_0-(1+2\xi)\ti{C}+2\xi J^{-1}-\la I\right)\,.
}
\end{array}
\eq
 Our aim is to show that the latter determinant equals $\la^{2M}$. Using
 the explicit form (\ref{q17}), (\ref{q17a}), (\ref{a6})
 of all entering matrices, one can verify that the first row of the
 matrix $C_0-(1+2\xi)\ti{C}+2\xi J^{-1}$
 consists of zeros. Also, $(2M-1)\times(2M-1)$ matrix obtained from
 $C_0-(1+2\xi)\ti{C}+2\xi J^{-1}$ by removing the first column and the first row is upper-triangular. Therefore,
\beq \label{a71}
\begin{array}{c}
\displaystyle{
 \det\limits_{2M\times 2M}\left(C_0-(1+2\xi)\ti{C}+2\xi J^{-1}-\la I\right)=\la^{2M}\,.
}
\end{array}
\eq
Together with (\ref{a3}) this
completes the proof for the $C_N$ root system and yields
\beq \label{a8}
\begin{array}{c}
\displaystyle{
\det\limits_{2N \times 2N}\left(\mathcal{L}^{1|1}\Big|_{BE (\ref{w34})}-\la I\right) = \la^{2N}\,.
}
\end{array}
\eq
The proof for the $D_N$ root system follows from the above at $\xi=-1$.

%%%%%%%%%%%%%%%%%%%%%%%%%%%%%%%%%%%%%%%%%%%%%%%%%%%%%%%%%%%%%%%%%%%%%%%%%%%%%%%%%%%%%%%%%%%%%%%%%%%%%%%%%%%%%%
%
\subsection{$C_N$ and $D_N$ root systems for ${\rm gl}(0|2)$ superalgebra}

Here $m=0$, $n=2$, $p(1)=1$, $p(2)=1$ and
 $g_1=0$, $g_2=\hbar$, $g_4=\sqrt{2}\hbar \, \xi$ for $C_N$ (and $\xi=0$ for $D_N$).

We are going to use the following
statement:

{\bf Lemma} {\em Consider $2N\times 2N$ matrix (\ref{w02}) for $C_N$ (and $D_N$) root system, i.e.
the Lax matrix $L(\{\dot q_j \}_N, \{q_j \}_N|\,g_1,g_2,g_4)$
 \beq\label{b11}
 \begin{array}{c}
 \displaystyle{
 L
 =\mats{ P+A}{ B}{- B}{-P-A}\,,\quad  A\,,B\in \Mat \,,
  }
  \end{array}
 \eq
where $P,A,B$ are given in (\ref{w03}). Then $\det(L-\lambda I)$ is an even
function of the parameter $g_4$
entering as common coefficient in the diagonal part of the matrix $B$.}

\noindent
The proof is given in the appendix.

Consider the Lax matrix
\beq \label{b2}
\begin{array}{c}
\displaystyle{
\mathcal{L}^{0|2} = L\left(\{H_j^{\rm{G}(0|2)}(\{q\}_N,\{\mu\}_M,\xi)\},\{q\}_N |\, 0,\hbar,\sqrt{2}\hbar \xi \right)\,,
}
\end{array}
\eq
where $H_j^{\rm{G}(0|2)}(\{q\}_N,\{\mu\}_M,\xi)$ are given in (\ref{w35}).
From the above Lemma we have
\beq \label{b3}
\begin{array}{c}
\displaystyle{
\det\limits_{2N \times 2N}\left (\mathcal{L}^{0|2}-\la I\right) =
\det\limits_{2N \times 2N} \left(L\left(\{H^{\rm{G}(0|2)}(\{q\}_N,\{\mu\}_M,\xi)\},
\{q\}_N |\, 0,\hbar,-\sqrt{2}\hbar \xi \right)-\la   I\right)\,.
}
\end{array}
\eq
Also notice that
\beq \label{b33}
\begin{array}{c}
\displaystyle{
H_{j}^{\rm G(0|2)}(\{z\}_N,\{\mu\}_M,\xi)=-H_{j}^{\rm G(2|0)}(\{z\}_N,\{\mu\}_M,-\xi)\,.
}
\end{array}
\eq
Therefore,
\beq \label{b4}
\begin{array}{c}
\displaystyle{
\det\limits_{2N \times 2N}\left (\mathcal{L}^{0|2}-\la I\right) =
\det\limits_{2N \times 2N} \left(L\left(\{H^{\rm{G}(0|2)}(\{q\}_N,\{\mu\}_M,\xi)\},
\{q\}_N |\, 0,\hbar,-\sqrt{2}\hbar \xi \right)-\la   I\right)
}
\\ \ \\
\displaystyle{
=\det\limits_{2N \times 2N} \left(-L^{T}\left(\{H^{\rm{G}(0|2)}(\{q\}_N,
\{\mu\}_M,\xi)\},\{q\}_N |\, 0,\hbar,-\sqrt{2}\hbar \xi \right)+\la   I\right)
}
\\ \ \\
\displaystyle{
 \stackrel{(\ref{b33})}{=}
\det\limits_{2N \times 2N} \left(L\left(\{H^{\rm{G}(2|0)}(\{q\}_N,\{\mu\}_M,-\xi)\},
\{q\}_N |\, 0,\hbar,-\sqrt{2}\hbar \xi \right)+\la   I\right).
}
\end{array}
\eq
 Finally, the Bethe equations (\ref{w36}) for ${\rm gl}(0|2)$ are exactly
 the same as in the ${\rm gl}(2|\,0)$ case (\ref{w32}) but with $\xi \rightarrow -\xi$.
 Thus the proof of the duality relation in this case follows from the one for ${\rm gl}(2|0)$
 (with sign of $\xi$ changed):
\beq \label{b5}
\begin{array}{c}
\displaystyle{
\det\limits_{2N \times 2N}\left(\mathcal{L}^{0|2}\Big|_{BE (\ref{w36})}-\la I
\right) = \la^{2N}\,.
}
\end{array}
\eq
%

%%%%%%%%%%%%%%%%%%%%%%%%%%%%%%%%%%%%%%%%%%%%%%%%%%%%%%%%%%%%%%%%%%%%%%%%%%%%%%%%%%%%%%%%%%%%%%%%%%%%%%%%
%%%%%%%%%%%%%%%%%%%%%%%%%%%%%%%%%%%%%%%%%%%%%%%%%%%%%%%%%%%%%%%%%%%%%%%%%%%%%%%%%%%%%%%%%%%%%%%%%%%%%%%%

\subsection{$B_N$ root system for ${\rm gl}(1|1)$ superalgebra}
Here $m=n=1$, $p(1)=0$, $p(2)=1$ and
$g_1=\sqrt{2}\hbar$, $g_2=\hbar$, $g_4=0$. The size of the Lax matrix is $(2N+1)\times(2N+1)$.

Introduce the matrix
\beq \label{c3}
\begin{array}{c}
\displaystyle{
\mathcal{L}^{1|1}_B = L\left(\{\tilde{H}_j^{\rm{G}(1|1)}(\{q\}_N,\{\mu\}_M)\},
\{q\}_N | \sqrt{2}\hbar,\hbar,0 \right)\,,
}
\end{array}
\eq
 where $\tilde{H}_j^{\rm{G}(1|1)}(\{q\}_N,\{\mu\}_M)$ are from (\ref{w041}).
 Notice that these eigenvalues
 coincide with those for ${\rm gl}(2|0)$ case (\ref{w039}), so that
\beq \label{c31}
\begin{array}{c}
\displaystyle{
\mathcal{L}^{1|1}_B = L\left(\{\tilde{H}_j^{\rm{G}(2|0)}(\{q\}_N,\{\mu\}_M)\},
\{q\}_N | \sqrt{2}\hbar,\hbar,0 \right)\,.
}
\end{array}
\eq
Then we use the determinant identity for the $B_N$ case (\ref{w50})
\beq \label{c4}
\begin{array}{c}
\displaystyle{
\det\limits_{(2N+1)\times(2N+1)}\left( \mathcal{L}^{1|1}_B -
\la I\right) = -\la^{2N-2M+1}\det\limits_{2M\times 2M}\left(\tilde{
\mathcal{L}}^{1|1}-\la I\right)\,,
}
\end{array}
\eq
 where
\beq \label{c33}
\begin{array}{c}
\displaystyle{
\tilde{\mathcal{L}}^{1|1} = L\left(\{-H_j^{\rm{G}(2|0)}(\{\mu\}_M,\{q\}_N,\xi=-1)\},\{\mu\}_M
 |\, 0,\hbar,\sqrt{2}\hbar \right)
}
\end{array}
\eq
is $2M\times 2M$ matrix
and eigenvalues $H_j^{\rm{G}(2|0)}(\{\mu\}_M,\{q\}_N,\xi=-1)$ are from
(\ref{w31}) for $C_N$ case with $\xi=-1$
and $\{q\}_N$, $\{\mu\}_M$ interchanged:
\beq\label{c35}
\begin{array}{c}
\displaystyle{
 H_i^{\rm{G}(2|0)}(\{\mu\}_M,\{q\}_N,\xi=-1)
 }
 \\ \ \\
 \displaystyle{
 =-\frac{\hbar}{\mu_i} +\sum\limits_{k\neq i}^{M}\left(\frac{\hbar}{\mu_i-
 \mu_k}+\frac{\hbar}{\mu_i+\mu_k} \right)
-\sum\limits_{l=1}^{N}\left(\frac{\hbar}{\mu_i-q_l}+\frac{\hbar}{\mu_i+q_l}\right)\,.
}
\end{array}
\eq
 Also, making the transposition we have:
\beq \label{c5}
\begin{array}{c}
\displaystyle{
\det\limits_{2M\times 2M}\left(\tilde{\mathcal{L}}^{1|1}-\la I\right) =
\det\limits_{2M\times 2M}\left(-(\tilde{\mathcal{L}}^{1|1})^{T}+\la I\right)
}
\\ \ \\
\displaystyle{
=\det\limits_{2M\times 2M}\left(L\left(  \{H_j^{\rm{G}(2|0)}(\{\mu\}_M,\{q\}_N,\xi=-1)\},
\{\mu\}_M |\, 0,\hbar,\sqrt{2}\hbar \right)+
\la I\right)\,.
}
\end{array}
\eq
The Bethe equations (\ref{w041}) imply that the last sum in (\ref{c35}) vanishes. Then
\beq \label{c51}
\begin{array}{c}
\displaystyle{
\det\limits_{2M\times 2M}\left(\tilde{\mathcal{L}}^{1|1}\Big|_{BE (\ref{w041})}-\la I\right)
}
\\ \ \\
\displaystyle{
 =\det\limits_{2M\times 2M}
 \left(L\left(  \{H_j^{\rm{G}(2|0)}(\{\mu\}_M,\{\emptyset\},\xi=-1)\},
 \{\mu\}_M |\, 0,\hbar,\sqrt{2}\hbar \right)+
\la I\right)\,.
}
\end{array}
\eq
In this way we come to the matrix $L\left(  \{H_j^{\rm{G}(2|0)}
(\{\mu\}_M,\{\emptyset\},\xi=-1)\},\{\mu\}_M |\, 0,\hbar,\sqrt{2}\hbar \right)$.
It is exactly the case (\ref{a4}), which was previously discussed in detail
 (\ref{a41})--(\ref{a71}) for generic $\xi$, and here we deal with $\xi=-1$. Therefore,
\beq \label{c52}
\begin{array}{c}
\displaystyle{
\det\limits_{2M\times 2M}\left(\tilde{\mathcal{L}}^{1|1}
\Big|_{BE (\ref{w041})}-\la I\right)=\lambda^{2M}\,.
}
\end{array}
\eq
Thus, plugging this into (\ref{c4}), we get
\beq \label{c6}
\begin{array}{c}
\displaystyle{
\det\limits_{2N \times 2N}\left(\mathcal{L}^{1|1}_B\Big|_{BE (\ref{w041})}-\la I\right)
= -\la^{2N-2M+1}\la^{2M} = -\la^{2N+1}\,.
}
\end{array}
\eq

\subsection{$B_N$ root system for ${\rm gl}(0|2)$ superalgebra}

Here $m=0$, $n=2$, $p(1)=1$, $p(2)=1$ and
$g_1=\sqrt{2}\hbar$, $g_2=\hbar$, $g_4=0$. The size of the Lax matrix is $(2N+1)\times(2N+1)$.

Introduce the Lax matrix
\beq \label{d3}
\begin{array}{c}
\displaystyle{
\mathcal{L}^{0|2}_B = L\left(\{\tilde{H}_j^{\rm{G}(0|2)}(\{q\}_N,
\{\mu\}_M)\},\{q\}_N | \sqrt{2}\hbar,\hbar,0 \right)\,,
}
\end{array}
\eq
where the eigenvalues $\tilde{H}_j^{\rm{G}(0|2)}(\{q\}_N,\{\mu\}_M)$ are from (\ref{w043}).
 Notice that
\beq \label{d33}
\begin{array}{c}
\displaystyle{
 \tilde{H}_j^{\rm{G}(0|2)}(\{q\}_N,\{\mu\}_M)=-\tilde{H}_j^{\rm{G}(2|0)}(\{q\}_N,\{\mu\}_M)
}
\end{array}
\eq
(see (\ref{w039})).
Therefore,
\beq \label{d4}
\begin{array}{c}
\displaystyle{
\det\limits_{(2N+1)\times(2N+1)}\left(\mathcal{L}^{0|2}_B -\la I\right) =
-\det\limits_{(2N+1)\times(2N+1)}\left(-\Big(\mathcal{L}^{0|2}_B\Big)^{T} +\la I\right)
}
\\ \ \\
\displaystyle{
=-\det\limits_{(2N+1)\times(2N+1)}\left( L\left(\{\tilde{H}^{\rm{G}(2|0)}
(\{q\}_N,\{\mu\}_M)\},\{q\}_N | \sqrt{2}\hbar,\hbar,0 \right) + \la I\right).
}
\end{array}
\eq
The Bethe equations for ${\rm gl}(0|2)$ case (\ref{w044})
are exactly the same as for the ${\rm gl}(2|0)$ case (\ref{w040}).
 Thus, the desired statement follows from the one for ${\rm gl}(2|0)$:
\beq \label{d5}
\det\limits_{(2N+1)\times(2N+1)}\left(\Big(\mathcal{L}^{0|2}_B\Big)\Big|_{BE (\ref{w044})}
-\la I\right) = -\la^{2N+1}\,.
\eq
%
%
%
%
%
%

%%%%%%%%%%%%%%%%%%%%%%%%%%%%%%%%%%%%%%%%%%%%%%%%%%%%%%%%%%%%%%%%%%%%%%%%%%%%%%%%%%%%%%%%%%%%%%%%%%%%%%%%%%%%%%%%%%%
%%%%%%%%%%%%%%%%%%%%%%%%%%%%%%%%%%%%%%%%%%%%%%%%%%%%%%%%%%%%%%%%%%%%%%%%%%%%%%%%%%%%%%%%%%%%%%%%%%%%%%%%%%%%%%%%%%%

\section{Conclusion}
\setcounter{equation}{0}

To summarize, let us formulate the final statement of the paper. Consider the set of supersymmetric
Gaudin models with boundary
based on the superalgebras ${\rm gl}(2|0)$, ${\rm gl}(1|1)$ and ${\rm gl}(0|2)$.
For each of the Gaudin model make the following substitutions into the data of the
classical Calogero-Moser
models of type $BCD$:
 \beq\label{w37}
    \displaystyle{
 z_j=q_j\,,\quad j=1\,, \ldots \,,N
 }
 \eq
 and
 \beq\label{w38}
    \displaystyle{
\dot q_j= H_j^{\hbox{\tiny{G}}}\quad\hbox{or}\quad \dot q_j=
{\widetilde H}_j^{\hbox{\tiny{G}}}\,,
 \ \ j=1\,, \ldots \,,N
 }
 \eq
 in the Lax matrix (\ref{w02}), which we denote as $L({\{\dot q_j \}}, {\{q_j \}}
 |\,g_1,g_2,g_4)$. Here $H_j^{\hbox{\tiny{G}}}$ and ${\widetilde H}_j^{\hbox{\tiny{G}}}$
 are eigenvalues of the Gaudin Hamiltonians (\ref{w29}) and (\ref{w037}) respectively.
 For the classical root systems the
 set of the coupling constants and the Lax matrix size $r$ are as follows:
 \beq\label{w87}
 \begin{array}{l}
\hbox{$B_N:$ ${\widetilde H}_j^{\hbox{\tiny{G}}}$
(\ref{w037}), $g_1=\sqrt{2}\hbar$, $g_2=\hbar$, $g_4=0$,
  $r=2N+1$;}
\\ \ \\
\hbox{$C_N:$  $H_j^{\hbox{\tiny{G}}}$  (\ref{w29}),
$g_1=0$, $g_2=\hbar$, $g_4=\sqrt{2}\hbar \, (\xi+p(2)-p(1))$,}
\hbox{ $r=2N$;}
\\ \ \\
\hbox{$D_N:$ $H_j^{\hbox{\tiny{G}}}$  (\ref{w29}) with
$\xi =p(1)-p(2)$, $g_1=0$, $g_2=\hbar$, $g_4=0$,}
\hbox{
$r=2N$.}
%\qquad\qquad\qquad\quad
 \end{array}
 \eq
If the Bethe roots $\{\mu_k\}$ satisfy the Bethe
equations (more precisely, (\ref{w038}) for the $B_N$ case,
(\ref{w30}) for the $C_N$ case and  (\ref{w30}) with $\xi =p(1)-p(2)$ for
the $D_N$ case), i.e.,
$H_j^{\hbox{\tiny{G}}}$ or ${\widetilde H}_j^{\hbox{\tiny{G}}}$ belong to the spectrum
of the Gaudin model,
then all eigenvalues of the Lax matrix
 $L({\{{\widetilde H}_j^{\hbox{\tiny{G}}}\}}, {\{q_j \}}|\,g_1,g_2,g_4)$ or
 $L({\{H_j^{\hbox{\tiny{G}}}\}}, {\{q_j \}}|\,g_1,g_2,g_4)$ and,
 therefore, all the integrals of motion, are equal to zero (see (\ref{q20})).
The extension of our results to ${\rm gl}(n|m)$ with arbitrary $n,m$ is a problem for
the future and is now under investigation.

%%%%%%%%%%%%%%%%%%%%%%%%%%%%%%%%%%%%%%%%%%%%%%%%%%%%%%%%%%%%%%%%%%%%%%%%%%%%%%%%%%%%%%%%%%%%%%%%%%%%%%%%%%%%%%%%%%%
%%%%%%%%%%%%%%%%%%%%%%%%%%%%%%%%%%%%%%%%%%%%%%%%%%%%%%%%%%%%%%%%%%%%%%%%%%%%%%%%%%%%%%%%%%%%%%%%%%%%%%%%%%%%%%%%%%%

\section{Appendix}
\def\theequation{A.\arabic{equation}}
\setcounter{equation}{0}

\subsection{Lax pairs and identities}

 The Lax matrices for the Calogero-Moser models (\ref{w01})
 are of sizes $(2N+1)\!\times\!(2N+1)$ (but they have effective size
 $2N\times2N$ when $g_1=0$) \cite{OP}:
 \begin{equation}\label{w02}
L =\left(
    \begin{array}{ccc}
    P+A & B & C     \\
    -B & -P-A & -C    \\
    -C^T & C^T & 0
        \end{array}
    \right)
    %,\qquad\ P,A,B\in{\rm Mat}_N
 \end{equation}
where $P,A,B$ are matrices of size $N\times N$ and $C$ is a column of
length $N$:
 \beq\label{w03}
  \begin{array}{c}
   \displaystyle{P_{ab} = {\dot q}_a}\delta_{ab}\,,\quad
  \displaystyle{A_{ab} = \frac{g_2 (1-\delta_{ab})}{q_a-q_b} }\,,
  \quad
%  \\ \ \\
  \displaystyle{B_{ab} = \frac{g_2 (1-\delta_{ab})}{q_a+q_b} +
  \frac{g_4\sqrt{2}\delta_{ab}}{2q_a}\,, \quad(C)_{a} =\frac{g_1}{q_a}\,,
  }
  %\quad\quad
  %\displaystyle{B_2 = -B_1
  %}\,,
  \end{array}
 \eq
 $a,b=1, \ldots ,N$. The Lax matrix
 provides
the Hamiltonians through $H_k=\frac{1}{2k}\, \tr L^k$.
For $k=2$ this yields (\ref{w01}). In fact, the matrix
(\ref{w02}) becomes the Lax matrix of the model (\ref{w01}) if
 the coupling constants $g_2$, $g_4$ and $g_1$ satisfy the
condition\footnote{The model (\ref{w01}) is integrable for arbitrary constants
but this Lax representation requires the constraint (\ref{w05}).
We use it since the factorization formulae and determinant
identities are available
for this type of the
Lax representation only. Alternative Lax pairs can be found in \cite{Feher}. }:
 \begin{equation}
 \label{w05}
  g_1(g_1^2 - 2g_2^2 + \sqrt{2}g_2g_4) = 0\,.
 \end{equation}
The classical root systems (of $BCD$ types) arise as follows:
 \begin{equation}
 \label{w06}
 \begin{array}{l}
\hbox{-- $B_N$ (${\rm so}_{2N+1}$): $g_4=0$, $g_1^2=2g_2^2$;
 $r=2N+1$, $x_{2N+1}=(q_1,...,q_N,-q_1,...,-q_N,0)$; }\hfill
\\ \ \\
\hbox{-- $C_N$ (${\rm sp}_{2N}$): $g_1=0$ and $r=2N$, $x_{2N}=
(q_1,...,q_N,-q_1,...,-q_N)$; }\hfill
\\ \ \\
\hbox{-- $D_N$ (${\rm so}_{2N}$): $g_1=0$, $g_4=0$
 and $r=2N$, $x_{2N}=(q_1,...,q_N,-q_1,...,-q_N)$,}\hfill
 \end{array}
 \end{equation}
 where  $r$ is equal to the size of the
 Lax matrix (it is a dimension of the fundamental representation)
  and $x_r=\{x_1,\ldots ,x_r\}$ is the set of coordinates on the Cartan subalgebra
 of the corresponding Lie algebra.

\paragraph{Factorization formulae.} Let us write down the factorized form
of the Lax matrices (\ref{w02})--(\ref{w06}) \cite{VZ}.
For this purpose we use the Vandermonde matrix $V$ and the
$D^0$ from (\ref{q16})--(\ref{q17}).
Both matrices are uniquely defined by a set of $r=N$ variables
$x_N=\{q_1,\ldots ,q_N\}$. We assume the following rule for
$BCD$ cases: the matrices $V,D^0,C_0,\ti C$ (\ref{q16})--(\ref{q17})
are of size $r\times r$ constructed by means of sets of variables
$x_r$ from (\ref{w06}). For $B_N$ root system $D^0$ obtained in this way
should be also multiplied by ${\rm diag}(I_N,I_N,\sqrt{2})$.
With these definitions, we have the following factorization formulae:

 ${\bf C}_{N}$ {\bf and} ${\bf D}_{N}$: set
 \begin{equation}
 \label{w101}
  \displaystyle{
 \hbox{$g_1=0$,\ $g_2=\hbar$,\ $g_4=\sqrt{2}\hbar\xi$}
 }
 \end{equation}
  and
 make the substitutions (these are some canonical transformations in the Hamiltonian approach)
 \begin{equation}
 \label{w10}
  \displaystyle{
  {\dot q}_i\rightarrow \frac{\xi\hbar}{q_{i}}+
  \sum\limits_{k\neq i}^{N}\Big(\frac{\hbar}{q_{i}-q_{k}}
+\frac{\hbar}{q_{i}+q_{k}}\Big)\,,\quad i=1, \ldots ,N\,.
 }
 \end{equation}
Then the matrix $L\rightarrow L'$ obtained in this way takes the form
 \begin{equation}\label{w11}
  \displaystyle{
 L'= \hbar (D^{0})^{-1}V( C_{0}-(1-2\xi)\tilde{C})V^{-1}D^{0}\,.
 }
 \end{equation}
This is true for the $C_N$ case, and $\xi =0$ in (\ref{w11}) yields the $D_N$ case.

  ${\bf B}_{N}$: set
 \begin{equation}
 \label{w141}
  \displaystyle{
  \hbox{$g_1=\sqrt{2}\hbar$, $g_2=\hbar$, $g_4=0$}
 }
 \end{equation}
  and
 make the substitutions
 \begin{equation}
 \label{w14}
  \displaystyle{
  {\dot q}_i\rightarrow \frac{2\hbar}{q_{i}}+\sum\limits_{k\neq i}^{N}\Big(\frac{\hbar}{q_{i}-q_{k}}
+\frac{\hbar}{q_{i}+q_{k}}\Big)\,,\quad i=1, \ldots ,N\,.
 }
 \end{equation}
 The matrix $L\rightarrow L''$ obtained in this way is represented in the form
 \begin{equation}
\label{w15}
  \displaystyle{
 L'' = \hbar (D^{0})^{-1}V(C_{0}+\tilde{C})V^{-1}D^{0},
 }
 \end{equation}
where $C_0$ and $\ti{C}$ are the matrices defined in (\ref{q17}), (\ref{q17a}) but of the
size $(2N+1)\times(2N+1)$.

%%%%%%%%%%%%%%%%%%%%%%%%%%%%%%%%%%%%%%%%%%%%%%%%%%%%%%%%%%%%%%%%%%%%%%%%%%%%%%%%%%%%%%%
%%%%%%%%%%%%%%%%%%%%%%%%%%%%%%%%%%%%%%%%%%%%%%%%%%%%%%%%%%%%%%%%%%%%%%%%%%%%%%%%%%%%%%%

\paragraph{Frobenius companion matrix}
\hspace{-3mm}
 %The companion Frobenius matrix
 is constructed by means of
  coefficients of characteristic polynomial $p(z) =
  \det(zI-J) = z^r+c_{r-1}z^{r-1}+c_{r-2}z^{r-2}+...+c_1z+c_0$.
  The matrix $J$ and its inverse are as follows:
\beq \label{w16}
J =\left( \begin{array}{ccccccc}
0 & 0 & .& . & .& 0 & -c_0\\
1 & 0 & .& .& .& 0 & -c_1\\
0 & 1 & .& .&. & 0 & -c_2\\
.&.&.&& &0&.\\
.&.& & .& &0&.\\
.&.& & & .&0&.\\
0&0&.&.&.&1&-c_{r-1}
\end{array} \right)\,,\quad
J^{-1} =\left( \begin{array}{ccccccc}
-c_1/c_0 & 1 & 0& . & .& . &0\\
-c_2/c_0 & 0 & 1& .& .& . & 0\\
-c_3/c_0 & 0 & 0& .&. & . & 0\\
.&.&.&& &.&0\\
.&.& & .& &.&0\\
.&.& & & .&.&1\\
-1/c_0&0&0&.&.&.&0
\end{array} \right)\,,
\eq
 where we assume
 generic case, so that $c_0\neq 0$. The zeros of $p(z)$ are eigenvalues
 $(x_1,\ldots ,x_r)$ of $r\times r$ matrix $J$, and
  $(-1)^i c_{r-i} =  e_i(x_1,\ldots ,x_r)$ are the elementary symmetric
  polynomials defined by $p(z)=\prod\limits_{k=1}^{r}(z-x_k) =
  \sum\limits_{k=0}^{r}(-1)^{r-k}e_{r-k}(x_1,..,x_r)z^k$.
  The ratios of coefficients entering $J^{-1}$ can be represented in the following way:
 \beq
 \label{w160}
   \displaystyle{
 \quad -\frac{c_k}{c_0}=(-1)^{k+1}\frac{e_{r-k}(x_1,\ldots ,x_r)}{e_r(x_1,\ldots ,x_r)}=(-1)^{k+1}e_k(x_1^{-1},\ldots ,x_r^{-1})\,.
  }
 \eq
The Vandermonde matrix $V_{ij}(x)=x_i^{j-1}$ brings $J$ to the diagonal form:
 \beq
 \label{w161}
   \displaystyle{
{\rm{diag}}(x_1,x_2,\ldots ,x_r) =VJ\,V^{-1}
  }
 \eq
or, equivalently,
 \beq
 \label{w162}
   \displaystyle{
{\rm{diag}}(x^{-1}_1,x^{-1}_2,\ldots ,x^{-1}_r) =VJ^{-1}V^{-1}.
  }
 \eq

%%%%%%%%%%%%%%%%%%%%%%%%%%%%%%%%%%%%%%%%%%%%%%%%%%%%%%%%%%%%%%%%%%%%%%%%%%%%%%%%%%%%%%%%%%%%%%%%%
%%
\paragraph{The determinant identities.}
For the $C_N$ and $D_N$
root systems, following \cite{VZ,VZZ}, consider the $2N\times 2N$ matrix
 \beq
 \label{w51}
 \displaystyle{
 {\mathcal
 L}=L\Big(\{{H}_j^{\rm G(2|0)}\}_N(\{q\}_N,\{\mu\}_M,\xi ),
 \{q\}_N|\,0,\hbar,\sqrt{2}\, \hbar \, \xi \Big)\,.
  }
 \eq
 It is the Lax matrix (\ref{w02}) $L({\{\dot q_j \}}, {\{q_j \}}
 |\,g_1,g_2,g_4)$ of type $C_N$, where velocities are replaced by eigenvalues
 of ${\rm gl}(2|0)$ Gaudin Hamiltonians (\ref{w31}), and the set of constants is chosen according to (\ref{w101}).
Define also the dual matrix $\widetilde {\cal L}$ of size $2M\times 2M$:
 \beq
 \label{w52}
 \displaystyle{
 \ti{\mathcal
 L}=L\Big(\{{H}_j^{\rm G(2|0)}\}_M(\{\mu\}_M,\{q\}_N,1-
 \xi ),\{\mu \}_M|\,0,\hbar,\sqrt{2}\, \hbar \, (1-\xi )\Big)\,.
  }
 \eq
Then the determinant identity for the matrices (\ref{w51}), (\ref{w52}) is as follows:
 \beq
 \label{w53}
 \displaystyle{
 \det\limits_{2N\times2N}\Big({\mathcal L}-\la I\Big)
  =\la^{2N-2M}\det\limits_{2M\times2M}\Big( \widetilde{\mathcal
  L}-\la I\Big)\,.
  }
 \eq

Similarly, for the $B_N$ root system define
the matrix
 \beq
 \label{w45}
 \displaystyle{
 {\mathcal
 L}=L\Big(\{\widetilde{H}_j^{\rm G(2|0)}\}_N,\{q_j \}_N|\,\sqrt{2}\hbar,\hbar,0\Big)
  }
 \eq
 of
size $(2N+1)\times(2N+1)$ with the set of coupling constants (\ref{w141}) and $\widetilde{H}_j^{\hbox{\tiny{G}}}$
are from (\ref{w039}).
 The dual matrix is of size $2M\times 2M$:
 \beq
 \label{w46}
 \displaystyle{
 \widetilde{\mathcal
 L}=L\Big(\{-{H}_j^{\rm G(2|0)}\}(\{\mu_j\}_M,\{q_j\}_N,\xi =-1),\{\mu_j\}_M|\,0,
 \hbar,\sqrt{2}\hbar\Big)\,.
  }
 \eq
Again, the arguments $\{q\}_N$ and $\{\mu\}_M$ are interchanged in
the expression (\ref{w46}):
 \beq
 \label{w47}
 \displaystyle{
 \widetilde{\mathcal L}
 =\mats{\ti A}{\ti B}{-\ti B}{-\ti A}\,,\quad  \ti A\,,\ti B\in \MatM \,,
  }
 \eq
 where
 \beq
 \label{w48}
 \displaystyle{
 {\ti A}_{ij}=\delta_{ij}\left(\frac{\hbar}{\mu_i} +
 \sum\limits_{k=1}^{N}\Big(\frac{\hbar}{\mu_i-q_k}+\frac{\hbar}{\mu_i+q_k}\Big)
  - \sum\limits_{l \neq i}^{M}\Big(\frac{\hbar}{\mu_i-\mu_l}+
  \frac{\hbar}{\mu_i+\mu_l}\Big)\right) + \frac{\hbar(1-\delta_{ij})}{\mu_i-\mu_j}
  }
 \eq
 and
 \beq
 \label{w49}
 \displaystyle{
 {\ti B}_{ij}=\delta_{ij}\frac{\hbar}{\mu_i}+(1-
\delta_{ij})\frac{\hbar}{\mu_i+\mu_j}\,.
  }
 \eq
Then the determinant identity reads as follows:
 \beq
 \label{w50}
 \displaystyle{
 \det\limits_{(2N+1)\times(2N+1)}\Big({\mathcal L}-\la I \Big)
  =-\la^{2N-2M+1}\det\limits_{2M\times2M}\Big( \widetilde{\mathcal
  L}-\la I \Big)\,.
  }
 \eq
 %

%%
%%

%%%%%%%%%%%%%%%%%%%%%%%%%%%%%%%%%%%%%%%%%%%%%%%%%%%%%%%%%%%%%%%%%%%%%%%%%%%%%%%%%%%%%%%%%%%%%%%%%

\subsection{The notation for ${\rm gl}(n|m)$ matrices}

We use the fundamental (defining)
representation of the ${\rm gl}(n|m)$ superalgebra.
Elements of ${\rm gl}(n|m)$ are endomorphisms of $\mZ_2$-graded vector space
$V = \mathbb{C}^{m|n}$.
The parity of the basis elements of $V$ is defined through $\mZ_2$-valued parameter
 \beq\label{w17}
p(i) = \left\{ \begin{array}{l}
0,\ \hbox{for}\ 1 \leq i \leq m,\\
1,\ \hbox{for}\ m+1 \leq i \leq n+m\,.
\end{array}\right.
 \eq
 The parity of the matrix units
 \beq\label{w171}
 \displaystyle{
p(E_{ij}) = p(i)+p(j)\ {\rm mod}\ 2
 }
\eq
 provides the rule for the tensor product of operators (matrices):
 \beq\label{w172}
 \displaystyle{
(A \otimes B)(C\otimes D) = (-1)^{p(B)p(C)}(AC)\otimes (BD)\,.
 }
\eq
The super-transposition for operator-valued matrices is defined as
 \beq \label{w23}
  \displaystyle{
A = \sum\limits_{i,j=1}^{m+n}E_{ij} \otimes a_{ij}
\quad\rightarrow\quad
A^t = \sum\limits_{i,j=1}^{m+n}(-1)^{p(j)+p(j)p(i)}E_{ji} \otimes a_{ij}\,.
 }
\eq
The super trace is
\beq \label{w24}
 \displaystyle{
\hbox{str} A = \sum\limits_{i=1}^{m}a_{ii} - \sum\limits_{i=m+1}^{m+n}a_{ii}\,.
 }
\eq
%

%%%%%%%%%%%%%%%%%%%%%%%%%%%%%%%%%%%%%%%%%%%%%%%%%%%%%%%%%%%%%%%%%%%%%%%%%%%%%%%%%%%%%%%%%%%%%%%%%%%%%%%%%%%%%%%%%%%

\subsection{Proof of Lemma (\ref{b11})}

Let us consider the
characteristic polynomial of the matrix $L$ (\ref{b11}):
\beq \label{b111}
 \displaystyle{
\det\limits_{2N \times 2N}
\left(L-\la I\right) = \det\limits_{N \times N}\Big( \left(B-A-P
\right)\left(B+A+P\right) + \la^2 I\Big)\,.
}
\eq
The explicit form of these matrices is as follows:
\beq  \label{b112}
\begin{array}{c}
\displaystyle{
\left(P+A+B\right)_{ij} = \delta_{ij}\left(\dot{q}_i+
\frac{g_4}{\sqrt{2}q_i} \right)+ \left(1-\delta_{ij}\right)\frac{2g_2q_i}{q_i^2-q_j^2},
}
\\ \ \\
\displaystyle{
\left(B-A-P\right)_{ij} = \delta_{ij}\left(\frac{g_4}{\sqrt{2}q_i}-
\dot{q}_i \right) - \left(1-\delta_{ij}\right)\frac{2g_2q_j}{q_i^2-q_j^2}.
}
\end{array}
\eq
 Next, compute the product of the matrices (\ref{b112}):
\beq \label{b113}
\begin{array}{c}
\displaystyle{
\sum\limits_{\al=1}^{N}\left(B-A-P\right)_{i\al}\left(B+A+P\right)_{\al i} =
\left(\frac{g_4^2}{2q_i^2}-\dot{q}_i^2 \right) +
\sum\limits^{N}_{\al \neq i}\frac{4g_2^2q_{\al}^2}{q_i^2-q_{\al}^2},
}
\\ \ \\
\displaystyle{
\sum\limits_{\al=1}^{N}\left(B-A-P\right)_{i\al}
\left(B+A+P\right)_{\al j} = \left(\frac{g_4}{\sqrt{2}q_i}-\dot{q}_i
\right)\frac{2g_2q_i}{q_i^2-q_j^2} - \left(\frac{g_4}{\sqrt{2}q_j}+\dot{q}_j
\right)\frac{2g_2q_j}{q_i^2-q_j^2}
}
\\ \ \\
\displaystyle{
- \sum\limits_{\al \neq i,j}^{N}\frac{4g_2^2q_{\al}^2}{(q_i^2-q_{\al}^2)(q_{\al}^2-q_j^2)}
= -\frac{2q_i \dot{q}_i}{q_i^2-q_j^2}-\frac{2q_j\dot{q}_j}{q_i^2-q_j^2}
-\sum\limits_{\al \neq i,j}^{N}\frac{4g_2^2q_{\al}^2}{(q_i^2-q_{\al}^2)
(q_{\al}^2-q_j^2)},\; i \neq j.
}
\end{array}
\eq
Thereby, we found that every matrix element of $N\times N$ matrix (\ref{b111}) is an even function of $g_4$. Thus the
characteristic polynomial (\ref{b111}) is an even function of $g_4$.

%%%%%%%%%%%%%%%%%%%%%%%%%%%%%%%%%%%%%%%%%%%%%%%%%%%%%%%%%%%%%%%%%%%%%%%%%%%%%%%%%%%%%%%%%%%%%%%%%%%%%%%%%%%%%%%%%%%
%%%%%%%%%%%%%%%%%%%%%%%%%%%%%%%%%%%%%%%%%%%%%%%%%%%%%%%%%%%%%%%%%%%%%%%%%%%%%%%%%%%%%%%%%%%%%%%%%%%%%%%%%%%%%%%%%%%

\section*{Acknowledgments}
\addcontentsline{toc}{section}{\hspace{6mm}Acknowledgments}

 The work
was supported in part by RFBR grants 18-01-00926 (M. Vasilyev and A. Zotov) and  18-01-00461 (A. Zabrodin).
The research of A. Zotov was also supported in part by the HSE University Basic Research Program, Russian Academic Excellence Project '5-100' and by the Young Russian Mathematics award.

%This work is supported by the Russian Science Foundation under grant 19-11-00062
%and performed in Steklov Mathematical Institute of Russian Academy of Sciences.

%The work of M. Vasilyev and A. Zotov
% (Sections 1-3.2)
% is supported by
%the Russian Science Foundation under grant 19-11-00062 and
% performed in Steklov Mathematical Institute of Russian Academy of Sciences.
% The work of A. Zabrodin
%  (Sections 3.3-5)
%  is supported by RFBR grant 18-01-00461.

%%%%%%%%%%%%%%%%%%%%%%%%%%%%%%%%%%%%%%%%%%%%%%%%%%%%%%%%%%%%%%%%%%%%%%%%%%%%%%%%%%%%%%%%%%%%%%%%%%%%%%%%%%%%%%%%%%%
%%%%%%%%%%%%%%%%%%%%%%%%%%%%%%%%%%%%%%%%%%%%%%%%%%%%%%%%%%%%%%%%%%%%%%%%%%%%%%%%%%%%%%%%%%%%%%%%%%%%%%%%%%%%%%%%%%%

\begin{small}

\end{small}

\end{document}